\documentclass[10pt,twocolumn,letterpaper]{article}

\usepackage[pagenumbers]{cvpr} 

\usepackage{graphicx}
\usepackage{amsmath}
\usepackage{amssymb}
\usepackage{booktabs}
\usepackage{paralist}
\usepackage{comment}

\usepackage[pagebackref,breaklinks,colorlinks]{hyperref}

\usepackage{multirow}
\usepackage{booktabs, tabularx}
\usepackage{comment}
\usepackage{arydshln}
\usepackage{xcolor}

\usepackage[capitalize]{cleveref}
\crefname{section}{Sec.}{Secs.}
\Crefname{section}{Section}{Sections}
\Crefname{table}{Table}{Tables}
\crefname{table}{Tab.}{Tabs.}

\usepackage{algorithm}
\usepackage{algorithmic}

\usepackage{comment}

\usepackage{tabularx}
\usepackage{multirow}
\usepackage{array}
\usepackage{amsmath}
\usepackage{bm}
\usepackage{colortbl}
\usepackage{caption}
\usepackage{placeins}
\usepackage{soul}
\usepackage{booktabs}
\usepackage{adjustbox}
\usepackage{nicematrix}

\begin{document}
\setlength\dashlinedash{0.2pt}
\setlength\dashlinegap{1.5pt}
\setlength\arrayrulewidth{0.3pt}
\title{Towards Consistent Language Models Using Declarative Constraints}

\author{Jasmin Mousavi\hspace{20pt}Arash Termehchy \\
Oregon State University\\
{\tt\small \{mousavij, termehca\}@oregonstate.edu}
}

\maketitle

\begin{abstract}
Large language models have shown unprecedented abilities in generating linguistically coherent and syntactically correct natural language output.
However, they often return incorrect and inconsistent answers to input questions.
Due to the complexity and uninterpretability of the internally learned representations, it is challenging to modify language models such that they provide correct and consistent results.
The data management community has developed various methods and tools for providing consistent answers over inconsistent datasets.
In these methods, users specify the desired properties of data in a domain in the form of high-level declarative constraints.
This approach has provided usable and scalable methods to delivering consistent information from inconsistent datasets.
We aim to build upon this success and leverage these methods to modify language models such that they deliver consistent and accurate results.
We investigate the challenges of using these ideas to obtain consistent and relevant answers from language models and report some preliminary empirical studies.  
\end{abstract}

\section{Introduction}
\label{sec:introduction}
Large language models (LLMs) have shown unprecedented abilities in processing natural languages \cite{radford2018improving,openai2023gpt4}. 
They effectively generalize to perform various tasks with few or no training examples.
Thus, there is a rapidly growing interest in using them to solve data-driven problems, such as, interactive question answering. 

Nonetheless, LLMs often provide incorrect answers to input queries and perform inaccurate inferences \cite{Ji_2023,openai2023gpt4}. 
Several studies indicate the recent LLMs provide {\bf up to 40\% erroneous answers to factual questions} \cite{openai2023gpt4}. 
These erroneous results are important obstacle for wide-spread use of LLMs in real-world applications. 

To address the problem of inaccurate answers returned by LLMs, we should recognize that {\bf LLMs are not knowledge bases, but rather approximate models of factual information}.
Due to this approximate nature, they may represent inaccurate and inconsistent patterns. 
They may over-generalize relationships in the pretraining data, which leads to returning spurious relationships and inaccurate results. 
The uninterpretable mixture of learned linguistic patterns and factual information has made it challenging to eliminate incorrect information from LLMs.

Nevertheless, we may be able to restrict LLMs' pre-trained representation or decoding to {\bf adhere to semantic constraints} in the domain to avoid generating incorrect results.
This is akin to the problem of {\bf data cleaning} and {\bf answering queries over inconsistent databases} \cite{AliceBook,Arenas:PODS:99:Consistent,bienvenu2013tractable,10.14778/1952376.1952378, DBLP:series/synthesis/2012Fan}.
Databases often contain data that does not comply with the semantic constraints in their domains.
For example, a person might not have any social security number or have more than one in a human resource database. 
The usual query processing methods might return inaccurate results over incomplete or inconsistent databases.
The data management community has developed a unified, usable, and scalable approach to repairing and querying inconsistent data based on {\bf declarative semantic constraints} \cite{Arenas:PODS:99:Consistent,10.5555/645505.656435,10.1145/3318464.3389708,Rekatsinas2017HoloCleanHD}.
Hence, instead of writing long and complex imperative programs, users specify the properties of the consistent dataset succinctly in high-level declarative languages. 
They are usually subsets of first order logic that are sufficiently expressive to capture important knowledge in the domain yet not too expressive to make reasoning intractable.  
Hence, data systems may check incompatibilities or redundancies in constraints efficiently. 
They can also be learned from data in an unsupervised manner \cite{papenbrock2015functional, baskaran2017efficient}.
This approach offers an end-to-end and unified method: data systems use these constraints both to clean or return reliable answers over inconsistent data.

There has been recent effort on limiting the decoded output of LLMs to follow some {\it syntactical patterns}, e.g., contain certain keywords \cite{lu-etal-2021-neurologic,lew2023sequential}.
In these systems, users often write (imperative) programs that detect some invalid patterns in the output of LLMs.
These systems, then, use constrained optimization or probabilistic inference over the sequences generated by the LLM to reduce the probability of the outputs with invalid patterns.
These efforts are steps in the right direction but fall short of providing a usable and scalable method to deliver consistent information over LLMs.
First, they do not generally support semantic constraints.
Second, users may have to write multiple and possibly long programs to {\it clean up} the output of the model.
As some domain may have numerous constraints, it is challenging to develop and maintain these programs.
Users must check manually whether these programs are consistent with each other and there is no redundancy across different programs.
Third, they are applied during the decoding stage, therefore, they can detect and eliminate only a limited set of inconsistencies.
Specifically, it is difficult for them to control all the implications of imprecise learned information in LLMs. 
For instance, the learned spurious relationships about one entity $e_1$ may impact how the LLM answers a question about a different but related entity $e_2$.
It is challenging to understand and apply implicit implications of imprecise information about $e_1$ when they analyze the output about $e_2$ during decoding.
Finally, modifying LLMs' outputs during decoding often reduces their (linguistic) coherency \cite{lu-etal-2021-neurologic,lew2023sequential}.


We believe that the success using of high-level semantic constraints in data management suggests that applying semantic constraints could similarly offer a practical and scalable approach for developing and maintaining reliable and consistent LLMs.
In this paper, we propose an end-to-end framework to provide a usable and unified approach to reduce inconsistencies in LLMs using high-level declarative constraints.
We leverage concepts and methods of data cleaning and querying or learning over inconsistent data in the data management community. 
We also investigate the challenges of using declarative constraints to pre-train accurate representations and infer precise answers from LLMs.
We discuss how to use current work on using declarative constraints to learn accurate ML models over inconsistent data \cite{picado2020learning,DBLP:conf/deem/ZhenCT23} and techniques to embed structured queries in vector space \cite{ijcai2019p845,DBLP:conf/iclr/RenHL20,jackermeier2023box2el} to address these problems. 
We also report preliminary results for integrating constraints in Llama-2 \cite{touvron2023llama}.

\section{Declarative Constraints} 
\label{sec:background}
The properties of entities and relationships in a domain of interest are often expressed as a set of (semantic) constraints.
Constraints take various forms depending on the data model used to represent information in the underlying domain. 
For example, in relational data, constraints are typically expressed as logical sentences involving relational predicates, such as functional dependencies \cite{AliceBook}.

Ontologies describe constraints on concepts, i.e., sets of entities, members of concepts (entities), and binary relationships between concepts (predicates) \cite{krotzsch2012description, 10.1145/2661643, ijcai2019p845, jackermeier2023box2el}. 
The simplicity of ontologies makes them a convenient choice for representing semantic properties of entities and relationships frequently found in textual data. 
Moreover, ontologies have been developed and are available across various domains, e.g., healthcare \cite{MedicalOntology23}.
Thus, in this paper, we assume that the semantic constraints are expressed in form of ontologies.
Nevertheless, we believe that the challenges and approaches discussed in this paper can be extended to other types of constraints.

To better understand this representation, let us consider an ontology for the domain of people. 
Concepts refer to sets of individuals, binary predicates denote relationships between people, and members of concepts represent individual names. For instance, the concept \textit{President} represents the set of presidents, while \textit{Democrat} represents  people registered as Democrat. Binary predicates like \textit{partyOf} establish relationships between individuals, indicating which presidents belong to the Democratic Party.

\section{Learning Consistent Representations}
Due to the lack of interoperability in a standard off-the-shelf LLM's learned representation, it becomes unclear whether these representations respect the semantic constraints of a given domain.
To address this problem, the LLM could be fine-tuned with the objective of learning a more consistent representation that encompasses semantic constraints.
%
%

The large number of parameters and intricate internal representations of LLMs make it difficult to directly enforce declarative constraints. 
The challenge lies in effectively incorporating these constraints into the pre-training or fine-tuning process without compromising the overall performance and generalization capabilities of the model.
To address this challenge, we need novel techniques that can include declarative constraints in an LLM learned representation. 

\subsection{Encoding Constraints in Input Sequences}
As declarative constraints are typically represented as ontologies, one approach to incorporating them into an LLM is by augmenting the context of the pre-training or fine-tuning input sequences. 
By including this as part of the input context, the LLM can leverage the additional information to generate semantically constrained responses.

To extract information from an ontology, it is necessary to first identify the entities that are relevant to the fine-tuning data. 
These selected entities then serve as a foundation for retrieving constraint information from the ontology.  
Incorporating information from an ontology into LLMs presents challenges, as LLMs are fine-tuned on unstructured data, while ontologies convey structured information. 
To overcome this, one may simply supplement the training data with information from the ontology in textual form, e.g., \textit{Obama is a President}. 
However, in domains containing several semantic constraints, the augmented training data may exceed the maximum sequence lengths (commonly restricted to 512 in most models). 

Another approach is to encode the ontology information into an embedded representation using an LSTM \cite{liu2022relational}. 
The embedding is integrated through the use of a gating function, allowing the language model to control what information augments the input. 
While this method successfully limits sequence length, it may not be optimal for incorporating constraints, as it could lead to information loss. 
Moreover, this method is more suitable for enriching the input with additional context rather than excluding incorrect information. 
Unfortunately, both these approaches fall short of incorporating the ontology in a manner that retains the structured information.

\subsection{Structural Embedding}
Ideally, the representation learned by an LLM should capture the structural information in the constraints.
Geometric embeddings (e.g. box, circle, cone) have been widely explored for learning representations of graph structures such as ontologies and knowledge bases \cite{xiong2023geometric, 10.5555/3495724.3497378, DBLP:conf/iclr/RenHL20, NEURIPS2019_cb12d7f9, ijcai2019p845, jackermeier2023box2el}. 
These embeddings aim to capture the inherent geometry and structure of the entities and relationships within the graph, including the set-based rules and logic constraints.
By constructing relational geometric embeddings, the resulting entity representations exhibit convex regions that not only model the similarity between entities but also encode set-based rules and logic constraints. 
These embeddings preserve the structural properties in an embedded space, ensuring that the output representations maintain the specified constraints.
For instance, if an ontology  has the constraint that \textit{every employee is a person}, the geometric embedding for \textit{person} should contain the geometric embedding for \textit{employee}, reflecting that employees are a subset of persons. 
The objective of learning geometric embeddings is to associate geometric points and concepts with geometric shapes in $\mathcal{R}^n$ such that the constraints of the ontology are preserved. 
Each constraint is associated with a loss function, and the overall objective is to minimize the total loss, which is the summation of the losses for each constraint.
These embeddings preserve the structural properties in an embedded space, ensuring that the output representations maintain the specified constraints.

Geometric embeddings have primarily been utilized in the context of ontology and knowledge bases \cite{ijcai2019p845, jackermeier2023box2el, 10.5555/3495724.3497378}. 
These applications focus on modeling structured data, while LLMs model unstructured data. 
It is not clear how to directly extend these concepts to LLMs.

Intuitively, the LLM should generate a probability distribution over a sequence of words in such a way that the probability of the next word, given the preceding context, lies within a predefined geometric shape that represents the desired constraints. 
To achieve this, we propose incorporating geometric embeddings into the LLM, enabling us to directly optimize the objective function with respect to the specified constraints. 
In addition to the traditional masked language modeling objective, we can introduce additional objective tasks that align with the constraints.
One such objective task is \textit{type modeling} \cite{parvez2018building}, which involves predicting the type of the next word given a set of ontology concepts $c$, a word $w$, and a type function $c(w)$ that indicates whether the word represents a concept entity (the word can also extend to a span of words). 
The goal is to predict the type of the next word based on this context.
Another objective task is \textit{entity composite modeling} \cite{parvez2018building}. 
In this task, given the previous context $\overline{w}$ and the type function $c(w)$, the objective is to predict the next word, which could be an entity name or any other relevant information related to the specified constraints.

During inference, the generated probabilities from the autoregressive language modeling, type modeling, and entity composite modeling tasks are combined to compute the final probability distribution over all possible words. 
This distribution represents the model's output and can be used to generate coherent and constraint-conserving responses.
By adding geometric embeddings and incorporating additional objective tasks, the model can be optimized to generate outputs that adhere to the desired constraints and exhibit meaningful behavior.

\section{Enforcing Constraints Using Prefix-Tuning}
Optimizing over multiple constraints can become computationally expensive, especially in the context of pretraining or fine-tuning of LLMs with numerous parameters. 
To address this challenge, one may employ a prefix-tuning technique, which offers a lightweight strategy for fine-tuning \cite{li2021prefix}.

Prefix-tuning involves freezing the gradients for most parameters during the fine-tuning process. 
It serves as a continuous method for tuning the prompt, which is otherwise discrete since it is represented in the word space. 
The core idea is to fine-tune the model for a specific task or constraint by adding a prefix vector to the input sequence. 
The prefix functions similar to prompt-tuning, effectively adding extra words to guide the model's behavior.

The key distinction is that the prefix is treated as virtual tokens that do not have associated parameters. 
The transformer can attend to the prefix as if it were actual tokens, allowing it to influence the model's behavior. The objective of prefix-tuning is to optimize the prefix specifically for the desired task or constraint. 
Since the gradients are turned off for the majority of the LLM, the computational overhead for prefix optimization is minimal.

By using prefix-tuning, an LLM can effectively support multiple tasks or constraints simultaneously. 
Each task can be associated with its own optimized prefix, allowing the model to adapt its behavior accordingly. 
This lightweight fine-tuning strategy allows the model to effectively model constraint-specific behavior while mitigating the computational complexity that arises when optimizing over a large number of constraints and parameters.

\section{Model Repair}
To ensure that a database complies with a constraint, we often find the information in the database that do not follow the constraint and update them so the database satisfies the constraint.
we may adopt this approach to repair a pretrained model so it satisfies a set of given constraints. 
In other words, we may find the portion of the model responsible for representing a constraint or lack thereof and update them if necessary so that the resulting model satisfies the constraint. 

\noindent{\bf Fact-based Repair.}
There has been some recent success in updating facts represented in an LLM by modifying its weights \cite{meng2023massediting}.
Each update aims at changing the object in a given triples in form of (subject ($s$), relation ($r$), object ($o$)). 
These methods find and modify the weights responsible for encoding $o$ so it represents the new object with high probability. 

Building upon this work, one may check whether a model satisfies a constraint by finding and modifying the pre-trained weights that represent the facts that violate the constraints.
We may represent each constraint as a set of facts that follow the constraint.

It is known that there are often many possible modifications of an inconsistent dataset to satisfy a set of constraints. 
It is challenging to maintain and query all these {\it repairs} of databases. 
Hence, researchers have proposed heuristics to choose a few of these repairs, e.g., the ones that differ the least from the original database.
The same problem might also happen in repairing models. 
One may use similar semantics and heuristics to maintain just a few repaired models.

\noindent{\bf Constraint-based Repair.}
It may take a long time to update a large number of facts in a model \cite{meng2023massediting}.  
Thus, the approach of fact-based repair may efficiently modify the model to satisfy constraints with a relatively few instances, e.g., facts in the ontology.
However, it might be computationally challenging to do for constraints with many instances.
Additionally, if a constraint has many instances, this approach might deliver many possible model repairs even after applying the aforementioned heuristics to reduce the space of possible repairs. 
It will be challenging to query or train these models for a given task.

LLMs generalize input data during pretraining. 
They have also been successfully used to generate data that closely resembles real-world data and train accurate models using a relatively few training examples for various tasks.
Hence, we hypothesize that they might represent some constraints in the domain in whole or in part.
If this hypothesis is true, an LLM does not satisfy some constraints because the LLM might represent them incompletely or erroneously. 
Hence, to ensure that the model satisfy a constraint, instead of repairing all facts that violate the constraint, one might change directly the portion of the model that represents a constraint.
This portion might be significantly smaller than the parts that represent the violating facts.
Thus, it might be substantially faster and easier to find the weights in the model responsible for incomplete or erroneous representation of the constraint than doing the same for all facts that violate that constraint.

\section{Consistent Querying and Prompting}
Since it is often resource-intensive to (re)train a large language model, users may like to use an available off-the-shelf language model.
Hence, they might not (re)train a large language model to apply semantic constraints. 
Additionally, as the learned representation in a language model contains a low-dimensional approximation of the training information, it is not clear whether it precisely and faithfully represent the relationships between entities and concepts according to the input constraints. 

To address this problem, one can use off-the-shelf language models and enforce the domain constraints during decoding.
A similar problem arises in querying inconsistent datasets.
Users often would like to avoid cleaning an inconsistent dataset due to its costs and overhead and prefer that the system returns answers to an input query that are consistent with the data constraints over the inconsistent dataset, i.e., {\it consistent query answering} \cite{Arenas:PODS:99:Consistent}. 
It has been shown that consistent answers can be computed efficiently for some classes of formal queries languages and constraints \cite{10.1145/3294052.3322190}.

There are, however, some challenges for adapting the idea of consistent query answering to return accurate results of language models.
First, consistent query answering is done via {\bf rewriting the input query} using the available constraints such that the resulting query encapsulates precisely the properties of answers to the query that adhere to the constraints.
The modified query is executed over the dataset and its answers are returned to the user.
As opposed to a database whose content is interpretable, it is challenging to interpret the learned representation of a language model.
Thus, it is not clear what the modified query will return over the language model. 

Second, questions or training information over language models are often in form of natural languages rather than formal query languages, such as SQL.
There have been some effort to develop formal query languages to prompt or ask questions from language models \cite{Beurer_Kellner_2023,MicrosoftLMQuery}.  
Nonetheless, users often would like to ask their questions in natural language. 
In what follows, we explain how to address these challenges.

\subsection{Query Rewriting As Chain-of-Thought Prompting}
Rewriting queries using constraints has generally two benefits for querying datasets.
First, the modified query may contain more information about the underlying domain, which are not available in the database \cite{10.1145/2661643}.
This may lead to more informative answers.
For example, assume that there are some domain constraints in form of an ontology.
One might be able to find new information about the relationships of entities, e.g., each patient is a human, which might not be explicitly available in the database.
Hence, the rewritten query might convey more information about the domain than the original one.
Second, the modified query expresses properties of the data items that are consistent with the domain constraints \cite{Arenas:PODS:99:Consistent,10.1145/3294052.3322190}.
Thus, it does not return inconsistent answers and improves precision of the original query.
There are algorithms to rewrite formal queries over relational databases in both cases. 
As we explained before, there are challenges of extending these ideas for querying language models.

Recently, researchers have observed that explaining the properties of the desired answers gradually, i.e., chain of thoughts, improves the accuracy of answering questions over language models \cite{wei2023chainofthought}.
One may use this property and provides the language model with step by step explanation of the modified query.
In particular, if the original query is written in a formal query language, such as SQL \cite{Beurer_Kellner_2023,MicrosoftLMQuery}, it lends itself to clear step by step explanation using its operators \cite{saeed2023querying}.  

If the original query is in a formal programming language \cite{Beurer_Kellner_2023,MicrosoftLMQuery,NormalLMQuery}, the current ideas in query rewriting using constraints might be used to modify the original query.
If the input query is in natural language, one should first identify the entities that appear in the query to find relevant constraints. 
Since constraints are usually written in subsets of first-order logic, one can express them in form of natural language, e.g., by translating $\subset$ to {\it is a subset of}. 
The final query will be a composition of the original one and natural language translation of its relevant constrains.



\subsection{Checking and Repairing Generated Sequences}
Due to the uninterpretablity of the learned representation in a language model, it is difficult to ensure that rewriting the query will return accurate answers.
Hence, one would also like to {\bf check} whether the returned answers from the language model comply with the constraints in the domain.
If the answers do not follow some constraints, we would like to {\bf repair and modify} them so that they adhere to the constraints while still being sufficiently relevant to the input question.

\subsubsection{Checking the Compliance of Output Sequences}
Data management community has provided several methods to check whether a structured dataset comply with a set of semantic constraints \cite{DBLP:series/synthesis/2012Fan,10.14778/1952376.1952378,10.1145/3294052.3322190,Rekatsinas2017HoloCleanHD}.
There are two important differences between checking the compliance of structured datasets, e.g., relational data, and sequences of tokens generated by a language model.
First, the output of popular language models is often in form of natural language to make it usable for end-users.
To check whether a sequence returned by a language model violates a constraint, one has to find tokens in the sequence that are members of concepts or mentions to entities or relationships in constraints.
This might be difficult as each entity might appear in different forms in a sequence of natural language tokens.
Second, checking the consistency of a dataset is often done offline and before query time.
Nonetheless, we have to check the generated sequences of language model during query time. 
Hence, it is essential to perform it very quickly.

As explained in Section~\ref{sec:introduction}, researchers have proposed methods that check the sequence of tokens generated by language models to satisfy lexical patterns \cite{lu-etal-2021-neurologic,ZhangICML23,lew2023sequential}.
In particular, authors in \cite{lew2023sequential} frame this as a probabilistic inference problem.
To answer a question, a language model may produce a set of candidate sequences and will return the one with largest probability of matching the input.
This method sequentially checks the output sequences as they are generated.
Given the probability of returning a sequence by the language model and the degree by which the sequence might violate the lexical patterns, this method computes posterior probability of satisfying the patterns for generated sequences and returns the one with the largest posterior probability.
This approach provides a clear semantic for the degree by which a sequence satisfies given patterns.

Nonetheless, it might be challenging to extend this approach for semantic constraints. 
Some domains may have many constraints, it might be too time-consuming to check to what degree a set of possible sequences satisfy all constraints in these domains during query time.
One might check constraints in order of their importance or probability of being violated based on previous observations to save some time.
We can also use current research on reasoning over constraints to find a minimal set of constraints that imply the entire set of reduce the number of constraints \cite{AliceBook}.

Another possible challenge is that one might have to check relatively long sequences of text to detect violations of semantic constraints. 
For example, the relationship between two entities might be represented in a relatively long sentence (paragraph) with each entity is placed in one end of the sentence (paragraph). 
The longer the size of examined sequences gets, the more generated sequences must be checked to compute the one with the largest posterior probability.
This may significantly increase the time of returning an answer to the user. 
To speed up this process, one might test the constraints in the order of how close the mentions to the concepts or relationships usually appear in the text to prune some candidate sequences early. 
Another useful technique is to consider a relatively small sample of possible sequences instead of the entire set to return the final consistent result.

\subsubsection{Semantic Constraint Detection in Text}
\label{sec:interaction}
To check whether a given sequence violates a constraint, we must identify in the sequence of generated tokens by the language model the mentions to the entities, concepts, or relationships in the constraint.
One might assume that there are knowledge bases that contains names of members of a concept or relationship in some domains \cite{liu2022relational}.
Also, natural language processing and information extraction communities have developed effective techniques to discover mentions to concepts, entities, or relationship in text \cite{zhou2022survey,pawar2017relation}.
These methods are often not exact and return a probability distribution over a set of possible concepts or relationships tokens represent.
One can consider this source of uncertainty during computing the posterior probability of satisfying constraints by the sequence. 

\subsubsection{Repairing Output Sequences}
If none of the output sequences satisfy the constraints to a sufficiently large degree, one might try to modify sequences so that they satisfy the domain constraints while still being relevant to the input question according to the language model.
The database community has proposed several approaches to repair an inconsistent structured dataset to comply with a set of semantic constraints efficiently \cite{DBLP:series/synthesis/2012Fan,10.14778/1952376.1952378,10.1145/3294052.3322190,Rekatsinas2017HoloCleanHD}.
These methods define a set of operations, e.g., insertion or deletion, to modify the input dataset so that the resulting dataset(s) satisfies the given constraints.
Since there are often various ways to repair an inconsistent dataset, researchers have proposed methods to return a reasonable subset of all possible repairs, e.g., that ones with minimal amounts of modifications.

One might extend the ideas to repair the decoded sequences of a language model by defining a set of operations, e.g, replacing or dropping tokens.
However, due to the complexity and heterogeneity of natural language, it is not clear whether we can precisely define the exact and generalizable conditions of applying a set of abstract operations under which the resulting sequence always satisfy the input constraints.
For example, the repair operations that make a sentence consistent with a given constraint might not do the same for another sentence.
It is also challenging to ensure that the repaired sequences are still syntactically correct, linguistically coherent, and relevant to the original question. 

To address this problem, one approach is to iteratively and incrementally modify the sequence in a greedy manner. After each modification, the resulting sequence can be checked to determine if it satisfies the input constraints.
Afterward, we might prompt the language model again with the repaired sequence and original question to return a linguistically coherent version of the sequence that are relevant to the question.
The prompt might explicitly ask the language model not to regenerate the original inconsistent sequence. 
One might also provide additional restrictions to the set of repair operations to increase the degree of relevance and coherency for the repaired sequence.  
To generate relevant repairs efficiently, we can generate the repairs for a small number of promising original sequences, e.g., the ones with highest posterior probabilities.







\begin{table*}[!ht]
\small
\centering

\begin{NiceTabular}{@{}lccccccccl@{}}[colortbl-like] 
\toprule
\multicolumn{1}{c}{\multirow{2}{*}[-2pt]{Method}} & \multicolumn{4}{c}{\textit{Generation Quality}} & \multicolumn{2}{c}{\textit{Constraint Satisfaction}} & \multicolumn{1}{c}{\textit{Efficiency}}
\\ \cmidrule(lr){2-5}\cmidrule(lr){6-7}\cmidrule{8-8}
           & ROUGE-L  & BLEU-4 & CIDEr    & SPICE  & Coverage & Satisfied & Time(s) \\
\specialrule{\lightrulewidth}{0pt}{0pt}
\rowcolor{gray!25}{ \it Prompt Only} &&&&&&&\\
ABS 0-shot & 25.37 & 06.29 & 04.34 & 13.81 & 51.61 & 18.84 &  01.56\\
ABS 1-shot & 29.46 & 08.41 & 06.22 & 18.49 & 74.49 & 35.34 & 01.85\\
ABS 2-shot & \textbf{31.34} & \textbf{10.60} & \textbf{07.33} & \textbf{20.06} & 76.74 & 38.74 & 01.83 \\
CNF 0-shot & 22.82 & 03.74 & 02.37 & 12.13 & 42.17 & 11.09 & 01.84 \\
CNF 1-shot & 29.60 & 08.07 & 05.88 & 18.77 & \textbf{77.47} & \textbf{38.88} & 01.51\\
CNF 2-shot & 30.93 & 09.92 & 06.81 & 19.22 & 74.48 & 34.34 & 01.46 \\

\rowcolor{gray!25} { \it Decoder Only} &&&&&&&\\
NL \cite{lu-etal-2021-neurologic}, beam=8 & 10.05 & 00.00 & 00.08 & 02.67 & 02.41  & 00.00 & 03.46  \\
NL \cite{lu-etal-2021-neurologic}, beam=32 & 10.36 & 00.00 & 00.06 & 02.31 & 01.12  & 00.00 & 12.47\\
NL \cite{lu-etal-2021-neurologic}, beam=64 & 9.74 & 00.23 & 00.04 & 02.59 & 00.96  & 00.00 & 24.01 \\
SMC \cite{lew2023sequential}, particle=8 & 23.10 & 02.60 & \textbf{01.71} & 15.37 & \textbf{100.0} & \textbf{100.0}  & 22.92\\
SMC \cite{lew2023sequential}, particle=16 & 22.86 & 02.52 & 01.62 & \textbf{15.55} & \textbf{100.0} & \textbf{100.0} & 22.96\\
SMC \cite{lew2023sequential}, particle=32 & \textbf{22.92} & \textbf{02.64} & 01.69 & 15.26 & \textbf{100.0}  & \textbf{100.0} & 23.17 \\

\rowcolor{gray!25}{ \it Prompt \& Decoder} &&&&&&&\\
ABS 0-shot + NL \cite{lu-etal-2021-neurologic}, beam=8 & 36.54 &14.82& 10.72 & 20.65  & 95.93 & 83.43 &  05.30  \\
ABS 1-shot + NL \cite{lu-etal-2021-neurologic}, beam=8 &39.07& 19.25 & 12.13 & 23.25 & 94.13 & 76.55 & 04.61   \\
ABS 2-shot + NL \cite{lu-etal-2021-neurologic}, beam=8 &39.39 & 19.76 & 12.26 & 23.65 & 93.81 & 75.48 & 05.11   \\
CNF 0-shot + NL \cite{lu-etal-2021-neurologic}, beam=8 &15.41 & 03.98 &01.56& 06.42 & 09.38 & 01.00 & 07.47   \\
CNF 1-shot + NL \cite{lu-etal-2021-neurologic}, beam=8 & 39.66 & \textbf{25.73} & \textbf{13.30} & \textbf{24.18} & 79.91 & 39.08 & 09.37   \\
CNF 2-shot + NL \cite{lu-etal-2021-neurologic}, beam=8 &\textbf{39.81} & 25.35 & 12.84 & 23.57 & 75.49 & 27.99 & 11.30   \\

ABS 0-shot + SMC \cite{lew2023sequential}, particle=8 &25.86 & 04.00 & 02.79 & 18.80 & \textbf{100.0} & \textbf{100.0} &  25.33  \\
ABS 1-shot + SMC \cite{lew2023sequential}, particle=8 &27.86 & 05.66 & 04.46 & 20.27 & \textbf{100.0} & \textbf{100.0} &25.14\\
ABS 2-shot + SMC \cite{lew2023sequential}, particle=8 & 28.62 & 06.17 &04.90& 20.55 & \textbf{100.0} & \textbf{100.0} &29.96\\
CNF 0-shot + SMC \cite{lew2023sequential}, particle=8 & 26.27 &04.07& 03.14 & 19.85 & \textbf{100.0} & \textbf{100.0} &  27.51  \\
CNF 1-shot + SMC \cite{lew2023sequential}, particle=8 & 27.40& 04.65 & 03.84 & 20.29 & \textbf{100.0} & \textbf{100.0} &  34.10  \\
CNF 2-shot + SMC \cite{lew2023sequential}, particle=8 &28.44 & 05.93  & 04.54 & 20.70 & \textbf{100.0} & \textbf{100.0} & 48.76  \\
\bottomrule
\end{NiceTabular}
\caption{Performance results on generation quality, constraint satisfaction, and efficiency over the CommonGen test set for different generation methods: \textit{decoder only}, \textit{prompt only}, and \textit{prompt + decoder}. Results include a comparison against soft constraint decoding, i.e., beam-based NeuroLogic (NL) \cite{lu-etal-2021-neurologic}, and hard constraint decoding, i.e., masked-based Sequential Monte Carlo (SMC) \cite{lew2023sequential}. Two prompting strategies were conducted: abstract based (ABS) and conjunctive normal form based (CNF). Each prompting strategy leveraged 0-shot, 1-shot, and 2-shot in-context examples. With the exception of efficiency (lower is better), a perfect score is 100. }
\label{tab:results}
\end{table*}

\section{Preliminary Empirical Results} 
\label{sec:experiments}
In this section we present our preliminary results for integrating lexical constraints with LLMs using the CommonGen benchmark \cite{lin2019commongen}. 
We identify the risks and trade-offs of augmenting LLMs with lexical constraints for the input and output layers, i.e., \textit{prompt only} and \textit{decoder only}, in terms of generation quality, constraint satisfaction, and efficiency. 
We also explore whether injecting constraints in multiple layers, i.e., \textit{prompt + decoder}, will help or hurt any risks and trade-offs that exist in single layer augmentation.  


\subsection{LLM Implementation Details}
We use Llama-2 \cite{touvron2023llama} as our pretrained language model across all experiments. 
Llama-2 was pretrained over 2 trillion tokens of data between January 2023 and July 2023.
Llama-2 consists of 7 billion parameters, 32 layers, 4096 hidden representation size, 32 attention heads, a 4096 token context window size.

\subsection{Dataset}
The CommonGen dataset \cite{lin2019commongen} is a benchmark designed for controlling language model generation with lexical constraints, i.e., contain certain keywords. 
Given a set of keywords, e.g., ``dog run field", the goal is to generate a sentence using all the keywords or the infections of the keywords, e.g., ``dogs" or ``dog".
Each set contains a minimum of 3 keywords and a maximum of 5 keywords.
The dataset is split into train, validation, and test sets of sizes 64.7k, 4.02k, and 1.5k rows, respectively. 
Typically, those using the CommonGen dataset would first fine-tune their language model using the training set. 
However, given the size of modern LLMs, users may not have the resources for fine-tuning an LLM. 
Hence, we focus on using inference-based algorithms that can be used with off-the-shelf models \textit{without fine-tuning}. Our results are conducted over the test set.

\medskip
\noindent \textbf{Lexical Constraints. }
In the CommonGen dataset, lexical constraints can be defined for a set of keywords $[w1, w2, w3]$ as follows.
If $S$ is a sentence, then $S$ must contain $w1$ or one of its inflections, $w2$ or one of its inflections, and $w3$ or one of its inflections. 
The objective here is to generate sentences that adhere to this constraint. 
A key characteristic of this constraint is its allowance for multiple valid outputs, stemming from the underspecificity of the input. 
This leads to a wide array of possible sentences that represent instances of the constraint.

\medskip
\noindent \textbf{Data Leakage. }
In recent years, researchers have become concerned with data leakage in LLMs \cite{carlini2021extracting,carlini2022quantifying}. Due to their ability to memorize training data \cite{carlini2022quantifying}, benchmark performance is often inflated. Given that Commongen is a public dataset, it is likely that Llama-2 \cite{touvron2023llama} has seen this dataset and even memorized ground truth sentences for the train and validation sets. However, since ground truth sentences for the test set are not publicly available, it is unlikely that Llama-2 memorized them.

\subsection{Input Layer Prompting Strategies}
We test two prompting techniques with varying representation sizes for lexical constraints on the CommonGen dataset \cite{lin2019commongen}. We also supply the prompt with additional in-context examples, i.e. \textit{0-shot}, \textit{1-shot}, and \textit{2-shot}. Examples were extracted from the training set.

\textbf{Conjunctive Normal Form (CNF)} prompting style models lexical constraints, i.e., keywords, in conjunctive normal form. For example, if the given concepts are ``dog run field", the lexical constraint in conjunctive normal form is (dog $\vee$ dogs $\vee$ ... ) $\wedge$ (run $\vee$ running $\vee$ ... ) $\wedge$ (field $\vee$ fields). We can translate this constraint to text by converting $\vee$ to \textit{or} and $\wedge$ to \textit{and}. Hence, our final prompt is ``Write a sentence using the words (dog or dogs or ... ) and (run or running or ... ) and (field or fields)".

\textbf{Abstract (ABS)} prompting style describes an abstract instance of a lexical constraint, e.g., ``Given a set of words $x$, write a sentence using all words in $x$ or inflections of $x$". Since \textit{ABS} prompts do not include specific instances of keyword inflections, it is more compressed than \textit{CNF} style prompts.

\subsection{Output Layer Decoding Strategies}
We test two decoding strategies with varying levels of satisfaction: soft constraint decoding with NeuroLogic \cite{lu-etal-2021-neurologic} and hard constraint decoding with Sequential Monte Carlo \cite{lew2023sequential}.

\textbf{NeuroLogic (NL)} \cite{lu-etal-2021-neurologic} is an inference time decoding algorithm that uses a variant of beam-search. The objective is to optimize the probability of generating sequences while also steering towards lexical constraints using a penalty term. 
Due to the interest of using off-the-shelf models, we chose \textit{not to fine-tune} an LLM for using the NeuroLogic decoder. 
It is important to note, however, their experiments were conducted using a fine-tuned model.

\textbf{Sequential Monte Carlo (SMC)} \cite{lew2023sequential} is an inference time masked decoding algorithm. They model sequence generation as a probabilistic inference problem using a variant of Sequential Monte Carlo with particle filtering. 
In SMC, a user writes a program that specifies the desired constraints in a sequential manner. 
The user may also specify the number of particles used, where each particle acts as a weighted sample of the posterior distribution. 
We programmed the task in CommonGen as an infilling problem, where keywords are sampled with a masked vocabulary.

\subsection{Metrics}
\textbf{Generation Quality} is measured using automatic metrics, such as ROUGE \cite{rouge}, BLEU \cite{bleu}, CIDEr \cite{cider}, and SPICE \cite{spice}. 
These metrics generate a quality score for the generated sentence based on human generated reference sentences, where a perfect score is 100. 

ROUGE-L is a precision and recall based metric that identifies the longest common co-occurring n-grams and sentence-level similarity by calculating the weighted harmonic mean.
BLEU-4 is a precision based metric that counts the matching 4-grams between the generated and reference sentences.
CIDEr is a consensus based metric that takes the average cosine similarity of Term Frequency - Inverse Document Frequency weighted n-grams.
SPICE is a semantic propositional based metric that establishes syntactic dependencies between words, then maps the syntactic dependencies using logical rules, and finally computes the F-score defined over the logical rules.

\textbf{Constraint Satisfaction} measures the method's ability to fully satisfy the constraint (generated sequence used all keywords or their inflections), i.e., \textit{satisfied}. We also calculate \textit{coverage}, which is an average over the percentage of keywords (or their inflections) used in the generated sequence.  

\textbf{Efficiency} is computed as the time taken (in seconds) for generating a sequence, i.e., inference time.

\subsection{Results \& Analysis}
Results over all experiments can be found on Table \ref{tab:results}. 

\subsubsection{Prompt Only }
We aim to understand how varying the constraint representation in the prompt, i.e., \textit{ABS} vs. \textit{CNF} and in-context examples, i.e., \textit{n-shot}, impact generation quality, constraint satisfaction, and efficiency. 

Across most experiments ABS prompting achieves higher satisfaction than \textit{CNF} prompting. This suggests that LLMs can understand abstract, high-level descriptions of constraints. Given the fact that \textit{CNF} prompts include all the inflections, one would expect higher constraint satisfaction across all experiments, however, this is not the case. With the exception of \textit{CNF 1-shot}, \textit{ABS} style prompting obtains higher satisfaction than \textit{CNF}. 

\textit{ABS} prompting outperforms \textit{CNF} prompting in terms of generation quality across all experiments.
\textit{CNF} style prompts are inherently more structured and further from `natural language' compared to \textit{ABS} style prompts. This suggests structured prompts are less beneficial and may require a fine-tuning strategy.

Increasing input length does not have significant impacts on efficiency. Despite \textit{ABS} prompts having a smaller constraint representation size than \textit{CNF} prompts, there is little change in inference time across all \textit{n-shot} experiments.

In-context examples boosts quality in both prompting strategies, but hurts satisfaction in \textit{CNF 2-shot}.
Including more than one in-context example worsens constraint satisfaction for \textit{CNF} style prompts. 
This suggests that extending the input context with inflections for every in-context example may lead to noisy, sub-optimal distributions during generation.

\subsubsection{Decoder Only}
In this section we discover the impacts of the output layer, i.e., decoder, on generation quality, constraint satisfaction, and efficiency. We compare two decoding strategies: soft constraint decoding, i.e., beam-based \textit{NL} and hard constraint decoding, i.e., masked-based \textit{SMC}.

In the absence of fine-tuning, beam-based/soft constraint decoding, i.e., NL, encounters challenges in both generation quality and constraint satisfaction. 
Compared with SMC, NL is more dependent on a high quality output distributions.
This suggest that soft constraint decoding may require higher quality output distributions from the LLM. 

Increasing the number of particles for SMC decoding does not yield quality or satisfaction improvements while increasing inference time.
This observation indicates that the underlying distribution may be of low quality, as increasing the number of particles does not enhance performance. Moreover, in cases of uncertainty, the decoder will not see benefits by increasing computational resources.

Although the SMC decoder achieves 100\% constraint satisfaction, this achievement comes at the cost of significantly reduced efficiency. 
For example, the longest inference time recorded among the tested prompting strategies was only 1.85 seconds, in contrast, SMC  with 8 particles required a considerably longer duration of 22.92 seconds for generation. This indicates a substantial increase in computational time required to achieve complete constraint satisfaction with masked decoding strategies, such as SMC.

Despite the improvements in constraint satisfaction, decoder only strategies tend to degrade generation quality and reduce efficiency.

\subsubsection{Prompt \& Decoder}
Although prompt only strategies have higher performance on generation quality and efficiency, they cannot provide any guarantees on constraint satisfaction. Conversely, decoding strategies optimize over constraint satisfaction, but at the cost of generation quality and efficiency. In this section we aim answer whether these two layers can work together to improve the disadvantages of single layer augmentation with a multi-layer approach, i.e., an end-to-end system. More specifically, we would like to understand how different strategies work together and whether they induce any trade-offs between our metrics.

Augmenting the prompt with constraints enhances generation quality and constraint satisfaction, indicating that prompting results in a higher-quality output distribution for the decoder to operates on. 
Notably, the \textit{NL} decoder, although underperforming as a standalone decoder, shows remarkable improvement in quality metrics when combined with prompts. This demonstrates that soft constraint decoding performance depends on the quality of the output distribution. 

Although prompting improves quality in \textit{SMC}, it has significant impacts on efficiency.
Checking for hard constraints within the \textit{SMC} decoding strategy is less scalable when compared to the implementation of soft constraints in \textit{NL}. 
In contrast, the \textit{NL} decoder benefits in both quality and efficiency.
Due to the higher quality output distributions produced with prompting, the \textit{NL} decoder spent less time searching, leading to reductions in inference time.

Across most experiments, the \textit{NL} decoder achieves higher generation quality than \textit{SMC}.
The use of soft constraints in \textit{NL} results in less drastic distribution changes compared to \textit{SMC}, allowing for higher quality generation, albeit with a trade-off in constraint satisfaction.

Despite the \textit{NL} decoder leveraging CNF formula, it exhibits higher satisfaction levels with \textit{ABS} style prompts. This indicates that structured prompts could potentially limit the model's performance by producing sub-optimal output distributions for the decoder. 
It suggests that high-level concepts and relationships might be more effective inputs to the model when optimizing the output distribution for decoding.

\section{Related Work}

{\bf Domain Specific Fine-Tuning.}
One can {\it fine-tune} an LLM on a set of domain-specific data sources to improve the quality of its answers for questions in a given domain \cite{10.1145/3366424.3383536}.
Nonetheless, it has shown that these methods may also lead to many inaccurate answers \cite{lu-etal-2021-neurologic}.
This is, in part, due to the fact that fine-tuning is inherently under-specified and may not sufficiently modify the model to eliminate its already learned spurious information.

{\bf Retrieval-Augmented LLMs.}
Researchers have augmented LLMs with {\it additional and potentially relevant information} from external data sources \cite{liu2022relational,borgeaud2022improving,mialon2023augmented}.
These methods often add extra information to the context considered during pretraining. 
This line of research have improved the accuracy of LLMs to a limited degree, as it does not address the core issue of having spurious and incorrect information in LLMs.
It is unclear whether adding more relevant information eliminate inaccurate information stored in the model.
Moreover, finding sufficiently many relevant data sources, particularly for long-tail entities, may pose challenges.
Compared to retrieval-based augmentation, we argue that constraints offer a more robust and adaptable framework for reducing inconsistencies in LLMs.
Constraints encapsulate rules governing the underlying domain, thereby enabling a system to generalize beyond particular instances in a dataset, i.e., out of distribution generalization.
They also are a generalization of retrieval-based approaches that augment LLMs with facts extracted from external sources, as each fact is a special case of a constraint.
Moreover, they extend beyond mere facts by representing more expressive relationships, such as instances where certain entities lack connections with other concepts. 
Constraints are also a form of high-level knowledge and effectively abstract large quantities of data. 
Their compressed representation offers a flexible and efficient method of augmenting LLMs by (1) allowing for soft incorporation of constraints (e.g. adhere to a constraint with 80\% probability), (2) reducing the size of information used as context to LLMs, and (3) providing a structured way to control the output of LLMs.

{\bf Semantic Parsing.}
LLMs have been an effective approach for program synthesis \cite{poesia2022synchromesh, singh2023format5}. 
These methods employ constrained decoding techniques to guarantee that the generated output aligns with the syntax and grammar of the target programming language. 
It is essential to note that the concept of ``semantic" in these contexts differs from our work.
In our work, semantic constraints serve as high-level rules that define relationships and characteristics between concepts.

{\bf Self-Consistency of Language Models.}
It is known that language models produce contradictory answers to the questions that seek the same information but phrased differently. 
Researchers have proposed methods to address this issue by prompting the language model to critique and refine its own output during inference \cite{madaan2023selfrefine}.
This method prompts the language model with differently phrased questions and builds a (weighted) model over answers to infer the most likely result.
We, however, mainly focus on ensuring that the language model follows semantic constraints. 

{\bf Extracting Knowledge from Language Models.}
Researchers have proposed methods to extract generic statements or factual knowledge from language models using prompt engineering and human supervision \cite{bhagavatula2023i2d2}.
The prompts are constructed in a way that encourages succinct factual statements.
They use human labeled data to detect inaccurate outputs and fine-tune the language model.
However, it might be challenging to collect a sufficient amount of training data to extract accurate statements.

{\bf Querying Language Models.}
There has been some recent effort to design programming languages for prompting large language models, i.e., {\it language model programming} \cite{Beurer_Kellner_2023,MicrosoftLMQuery,NormalLMQuery}. 
There are generally domain-specific programming languages to extract information from and control the output of a large language model to satisfy the users' input hard constraints, akin to where conditions in SQL queries. 
Some of these languages resemble database query languages, e.g., SQL \cite{MicrosoftLMQuery}.
These languages aim at making it easier to query and prompt and optimize the number of calls to large language models.
However, these languages do not generate consistent results conditioned on domain constraints.
Thus, they may return answers that violate semantic constraints in the domain. 

\bibliographystyle{ieee_fullname}
\bibliography{main}

\end{document}